\documentclass[twocolumn,aps,nofootinbib,prl,groupedaddress]{revtex4}
\usepackage{epsfig,amssymb,amsmath}
\usepackage[hypertex,linkcolor=red]{hyperref}
\def\comment#1{}
\def\section#1{{\par\em #1:--- }}
\def\togli#1{}

\def\>{\rangle}
\begin{document}
\title{A reply to ``Problems with modelling closed timelike curves as
  post-selected teleportation''} \author{Seth Lloyd$^{1}$, Lorenzo
  Maccone$^{1,2}$, Raul Garcia-Patron$^{1,3}$, Vittorio
  Giovannetti$^{4}$, Yutaka Shikano$^{1,5}$, Stefano Pirandola${^6}$,
  Lee A.  Rozema$^{7}$, Ardavan Darabi$^{7}$, Yasaman
  Soudagar$^{7,8}$, Lynden K.  Shalm$^{7}$, and Aephraim M.
  Steinberg$^{7}$} \affiliation{$^{1}$xQIT,Massachusetts Institute of
  Technology, 77 Mass Ave, Cambridge MA.\\
  $^2$Dip. di Fisica ``A. Volta'' and INFN Sez. Pavia, via Bassi 6,
  Pavia, Italy\\
  $^3$Max-Planck Institut f\"ur Quantenoptik, Hans-Kopfermann?Str. 1,
  D-85748 Garching, Germany\\
  $^4$NEST-CNR-INFM \& Scuola Normale Superiore, Piazza dei Cavalieri
  7, I-56126, Pisa, Italy. \\ $^5$Dep. Physics, Tokyo Institute of
  Technology, 2-12-1 Oh-Okayama, Meguro, Tokyo, 152-8551, Japan.\\
  $^6$ Department of Computer Science, University of York, York YO10
  5GH, UK.  \\ $^7$CQIQC,IOS, Department of Physics, University of
  Toronto, Canada M5S 1A7.  \\$^8$Laboratoire des fibres optiques,
  \'Ecole Polytechnique de Montr\'eal, Eng. Phys. Dep., Montr\'eal,
  Canada.}

\begin{abstract}
  In arXiv:1107.4675 Ralph uses our post-selection model of closed
  timelike curves (P-CTC) to construct an ``unproven-theorem''
  paradox, and claims that this voids our argument that P-CTCs are
  able to resolve such types of paradoxes. Here we point out that
  Ralph has not accounted for all the interactions needed for his
  construction.  A more careful analysis confirms that indeed there is
  no paradox, contrary to his claims.
\end{abstract}
\maketitle

In \cite{ctc} we proposed a quantum prescription for dealing with the
closed timelike curves that arise (for example) from certain solutions
of Einstein's equations of general relativity \cite{GODEL}. Our model,
P-CTC, is based on quantum teleportation with post-selection. Closed
timelike curves allow for non-causal trajectories in spacetime, also
known as time machines. Such devices notoriously generate time-travel
paradoxes, such as the ``unproven-theorem'' paradox: a time traveller
reads a theorem in a book, travels back in time and narrates this
theorem to a mathematician of the past that writes the theorem in a
book, {\em the same book the traveller will consult}. While not a
logical contradiction, this paradox is inherently unsatisfactory
\cite{DEU}.  In \cite{ctc} we gave an argument on how P-CTC might
resolve the unproven theorem paradox.

\begin{figure}[htb]
\begin{center}
\epsfxsize=.9\hsize\epsffile{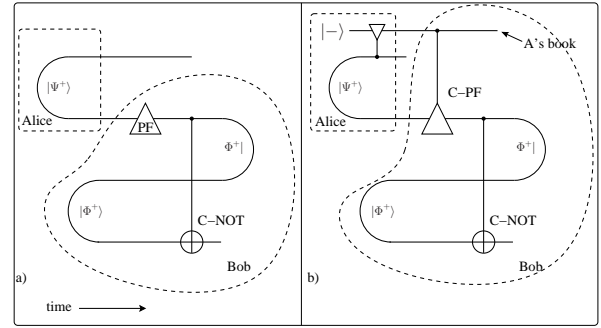} 
\end{center}
\caption{ Ralph's argument and its completion (time flows
  left-to-right in this diagram). a) Ralph's argument.  The $\subset$
  symbols denote a maximally entangled state:
  $|\Psi^+\>=(|10\>+|01\>)/\sqrt{2}$ for Alice and
  $|\Phi^+\>=(|00\>+|11\>)/\sqrt{2}$ for Bob, the $\supset$ symbol
  denotes a projection onto $|\Phi^+\>$. Bob mans the phase flip (PF)
  and sends information to Alice to his past, by using the C-NOT based
  circuit depicted. b) The completion of Ralph's argument: a
  purification of Bob's action (reading of Alice's book and writing
  the same information to the PF) must be done in terms of a
  controlled-unitary (C-PF) which is controlled by Alice's book (or by
  some environmental degree of freedom correlated to Alice's book).
  This completed circuit leads to no paradox.\label{f:utp}}
\end{figure}

In \cite{ralph} Ralph has employed our model to purportedly generate
an unproved-theorem paradox, claiming that this voids our argument
above. In particular, Ralph uses a well known nonlinear mechanism of
CTCs \cite{looplong} to allow Bob to send information to the past
using a phase flip transformation (see Fig.\ref{f:utp}a). This by
itself is not sufficient to generate paradoxes, since it is well known
that self-consistent time loops are possible (e.g.~see
\cite{billiards}). However, Ralph suggests that Alice can use the
information that Bob is sending back in time to her to write a theorem
in a book. In Alice's future, Bob uses the same book on which Alice
has written the theorem to decide which information to send her back
in time. The theorem has come out of nowhere since Alice has learned
it from Bob, and Bob from Alice: the unproven-theorem paradox.

A more careful analysis shows that Ralph's argument is incomplete,
since he has not analyzed the transformation that Bob must implement
to read the theorem from Alice's book and to write this information
into the CTC through the phase-flip transformation (to send it back in
time to Alice). When one analyzes the action of such transformation,
then it becomes clear that the post-selected teleportation mechanism
of the P-CTC intervenes and prevents the paradox from happening.

To read Alice's book and to write it into the CTC with the phase-flip,
Bob will use a unitary transformation: in fact, any physical
transformation can always be purified into a unitary, at the expense
of considering also the evolution of the environment. In other words,
Bob's phase-flip must be part of a controlled-unitary that uses as
input Alice's book, or something (e.g. some part of the environment)
correlated to Alice's book (see Fig.\ref{f:utp}b).

Let us analyze two possible controlled unitaries that Bob can use to
implement Ralph's proposal. The first one is a simple C-NOT in the
$+,-$ basis, where the control bit is Alice's book (or some
environmental degree of freedom correlated to Alice's book) and the
controlled bit is the one he sends through the CTC, Ralph's phase-flip
transformation. The C-NOT acts as follows on a basis: $|++\> \to
|+-\>\;;\ |+-\> \to |++\>\;;\ |-+\> \to |-+\>\;;\ |--\> \to |--\>$.
Implementing Bob's transformation in this way, we obtain an
impossibility when post-selection is introduced: the amplitude for the
whole process of Fig.\ref{f:utp}b is always zero. The second one is a
``copy operation on the $+,-$ basis'', where the first bit (Alice's
book) is the source, and the second bit (Bob's phase-flip) is the
destination, namely $|++\>|+\>_e \to |++\>|+\>_e\;;\ |+-\>|+\>_e \to
|++\>|-\>_e\;;\ |-+\>|+\>_e \to |--\>|-\>_e\;;\ |--\>|+\>_e \to
|--\>|+\>_e$\;; {\em etc.}, where an environmental qubit $e$ has been
added to guarantee unitarity. Implementing Bob's transformation in
this way (initializing the environmental qubit in $|+\>_e$), we find a
tautology: Bob can only write a single fixed state ($|+\>$) as his
theorem, namely his theorem is vacuous. In fact, one cannot argue in
this case that the state $|+\>$ is the theorem itself: it is true that
the appearance of this state is determined by the structure itself of
the P-CTC (i.e.~Alice's state and Bob's actions).  However, once this
has been established, the map has a single pure-state fixed point that
cannot sustain arbitrary information that could encode an {\em
  arbitrary} theorem (in contrast, for example, to Deutsch's
map~\cite{DEU}). Even though both Alice and Bob can choose which will
be the single fixed point of the map, this is inequivalent to the
unproven-theorem paradox (the theorem appearing out of nowhere).  On
the contrary, if Bob is aware of Alice's choice of initial state, he
can write a theorem in the time loop with his choice of
transformations.  But that implies that Bob is the theorem's author:
it has not appeared out of nowhere.  [Equivalently, if Alice is aware
of what transformations Bob will implement, she can write the theorem
in the loop with her choice of the initial state, but again this is
not an unproven-theorem paradox because she definitely is the author
herself.] Analogous considerations will hold for physical realizations
of the P-CTC different from the one analyzed by Ralph.

Note that in \cite{ctc} we give a completely different resolution to
the unproven-theorem paradox. 

In conclusion, Ralph's treatment is incomplete so that he fails in
obtaining an unproved-theorem paradox from his scheme, contrary to his
claims.


\begin{references}
\bibitem{ctc} S.Lloyd, et. al, Phys. Rev.  Lett. {\bf 106}, 040403
  (2011).
\bibitem{GODEL} K. G\"{o}del, Rev.  Mod. Phys. {\bf 21}, 447 (1949).
\bibitem{DEU}D. Deutsch, Phys. Rev. D {\bf 44}, 3197 (1991).
\bibitem{ralph} T. Ralph, preprint arXiv:1107.4675 (2011).
\bibitem{looplong} e.g. see J.B. Hartle, Phys. Rev. D {\bf 49}, 6543
  (1994); H.D. Politzer, Phys. Rev. D {\bf 49,} 3981 (1994); S.
  Rosenberg, Phys. Rev. D {\bf 57}, 3365 (1998); S. Lloyd, et al.
  Phys. Rev. D {\bf 84}, 025007 (2011).
\bibitem{billiards}F. Echeverria, G. Klinkhammer, and K.S. Thorne,
  Phys. Rev. D {\bf 44}, 1077 (1991).
\end{references}
\end{document}